\documentclass[sigconf]{acmart}
\AtBeginDocument{%
  \providecommand\BibTeX{{%
    \normalfont B\kern-0.5em{\scshape i\kern-0.25em b}\kern-0.8em\TeX}}}

\acmConference[MSR 2022]{MSR '22: Proceedings of the 19th International Conference on Mining Software Repositories}{May 23–24, 2022}{Pittsburgh, PA, USA}

\usepackage{algorithmic}
\usepackage{graphicx}
\usepackage{textcomp}
\usepackage{xcolor}
\usepackage{url}
\usepackage{listings,tikz,beramono}
\usepackage{multirow}
\usepackage{textcomp} 
\usepackage{enumitem} 
\usepackage{float}
\def\BibTeX{{\rm B\kern-.05em{\sc i\kern-.025em b}\kern-.08em
    T\kern-.1667em\lower.7ex\hbox{E}\kern-.125emX}}
\usepackage{xspace}
\newcommand*{\eg}{\emph{e.g.,}\@\xspace}
\newcommand*{\ie}{\emph{i.e.,}\@\xspace}
\newcommand*{\etal}{\emph{et al.}\@\xspace}
\usepackage{booktabs}
\definecolor{dark green}{rgb}{0.0, 0.7, 0.0}
\usepackage{tcolorbox}

\newlength\MAX  

\definecolor{MyPink}{HTML}{FF00FF}

\usepackage[export]{adjustbox}
\include{CodeOverlayMagic}

\begin{document}
\graphicspath{{images/}}

\title{Characterizing High-Quality Test Methods:\\A First Empirical Study}



\author{Victor Veloso, Andre Hora}
\affiliation{%
  \institution{\emph{Department of Computer Science} \\ Universidade Federal de Minas Gerais (UFMG)}
  \city{Belo Horizonte}
  \country{Brazil}}
\email{{victorveloso,andrehora}@dcc.ufmg.br}

\renewcommand{\shortauthors}{Veloso and Hora}

\begin{abstract}
    To assess the quality of a test suite, one can rely on mutation testing, which computes whether the overall test cases are adequately exercising the covered lines. 
    However, this high level of granularity may overshadow the quality of individual test methods.
    In this paper, we propose an empirical study to assess the quality of test methods by relying on mutation testing at the method level.
    We find no major differences between high-quality and low-quality test methods in terms of size, number of asserts, and modifications.
    In contrast, high-quality test methods are less affected by critical test smells.
    Finally, we discuss practical implications for researchers and practitioners.
\end{abstract}



\maketitle

\section{Introduction}

Software testing is a key practice in software development.
As a proxy of test quality, one can rely on code coverage or mutation analysis.
{Coverage measures the percentage of code that is covered by tests and is typically used to assess the test effectiveness~\cite{hilton2018large, ivankovic2019code, coverage_py}.}
{ However, it presents some limitations~\cite{coverage_best_practices, fuzzingbook2019_index, 10.1145/2896941.2896944}.}
{For example, one can have great coverage with no assert~\cite{fuzzingbook2019_index}.} 
Another solution to assess test quality (and overcome such limitations) is mutation testing~\cite{tse_mutation_2011, coverage_best_practices, fuzzingbook2019_index, pitest, coles2016pit}.
{This technique injects mutations (artificial faults) into the code and checks if tests can detect (or ``kill'', in the mutation testing terminology) these mutations.
The rationale is that if it fails to detect such mutations, it will miss real bugs~\cite{fuzzingbook2019_index}.}

Code coverage and mutation score typically target the overall test suite effectiveness~\cite{codecov, coveralls, coverage_py, pitest}.
However, this high level of granularity may overshadow test quality~\cite{hilton2018large}. 
Mutation Testing at the method-level has proven useful for test suite reduction~\cite{10.1145/2635868.2635921} and assessing test quality with it would be important to:
(1) understand the characteristics of good (and bad) test methods,
(2) provide novel empirical data at method level to differentiate both good and bad test methods, and
(3) help to improve the quality of existing tests.

In this paper, we propose an empirical study to assess the quality of \emph{test methods} by relying on mutation testing.
We extend a state-of-the-art mutation testing tool~\cite{pitest} to analyze test methods and report mutation results at the method level.
Then, we assess 18,321 test methods provided by five popular open-source projects: RxJava, OkHttp, Retrofit, ZXing, and Apache Commons Lang.
We then propose research questions to assess high-quality test methods:
\begin{enumerate}[label=\emph{RQ\arabic*:},start=1,topsep=0px,partopsep=0px]
    
    
    
    \item \emph{What are the code and evolutionary characteristics of high-quality test methods?}

    
    \item \emph{What test smells are prevalent in high-quality test methods?}
    
    
\end{enumerate}

Overall, we find no major differences between high and low-quality test methods in terms of size, number of asserts, and modifications.
In contrast, high-quality test methods are less affected by critical test smells.

\noindent\emph{Contributions:}
The contributions of this paper are twofold:
(1) we provide an empirical study to characterize high-quality test methods and
(2) we discuss implications for practitioners and researchers working on software testing.

\smallskip

\noindent Dataset: \url{zenodo.org/record/4987677#.YMz8G5pKiA0}

\section{Mutation Testing In a Nutshell}
\label{sec:background}



{Test mutation technique assesses test effectiveness in four major steps (illustrated in Figure~\ref{fig:OriginalPITFlow}-left).}
First, the project test suite is executed and the results are stored as the \emph{expected output}.
Then, a mutation testing engine (\eg~PIT~\cite{pitest}) parses the code and applies mutation operators on code structures generating a set of mutants.
The mutants are separately tested by the test suite and the results form the \emph{obtained output}.
Lastly, each \emph{obtained output} is compared to the \emph{expected output}.
{A mutant is ``killed'' when at least one of the test results differs between both sets, \ie when at least one of the test methods run on the mutants failed, meaning they properly detected the code mutations.}
{Finally, a mutation score is computed: higher scores mean the test suite is better in catching real bugs~\cite{fuzzingbook2019_index}.}

\subsection{Mutation Score Computation}\label{sec:ScoreComp}

The mutation score is defined as the ratio of killed mutants and the number of generated mutants {(which includes the \emph{killed}, \emph{survived}, and \emph{uncovered} sets)}. 
%
%
A mutant is \emph{killed} in three scenarios: \emph{failure}, \emph{error}, or \emph{time-out}.
A \emph{failure} happens when the test fails, \ie an assertion detected the modification.
{An \emph{error} occurs when an exception is raised.}
Lastly, a \emph{time-out} happens when the test execution takes considerably longer, possibly leading to an infinite loop. 
%
%
{\emph{Survived} happens when the test passes, \ie no assertion detected the modification.}
{Uncovered mutations cannot be killed, because no test run reached them, hence PIT skips their execution.}
%
%

Figure~\ref{fig:ExampleTable} exemplifies the execution of mutation testing based on three mutation operators.
The target system has a production class (\textit{SUT}) with two methods, \textit{sum()} and \textit{triangle()}.
{It also has nine test methods: three cover \textit{sum()} and six cover \textit{triangle()}.}
{For simplicity, we do not show the code of the test methods, but their asserts (column ``Assertion'').}
{The four generated mutants are annotated in the \textit{SUT} source code and detailed in the boxes.}
{For example, Mutation 1 replaces the ``+'' (sum) operator with ``-'' (subtraction).}
{Also, for each test method, Figure~\ref{fig:ExampleTable} shows its related mutants, the obtained result, and the status of the mutant.}
{At last, the mutation score for the target system is 100\% because} 
{ \textit{testSum1()} kills mutants 1 and 2, while \textit{testTriangle1()} kills mutants 3 and 4.}


\subsection{Limitation of Test Suite Mutation Testing}

Although test suite mutation testing is ideal for gathering the overall test quality in a system, it has three limitations:
{(1) the overall system mutation score overshadows the quality of individual test methods;}
{(2) the quality of a contribution (\eg~a pull request with code and tests) can be unnoticed in a large system because its score may be unaffected by small code changes (due to existing tests outnumber the contributed tests);}
{(3) existing test methods may kill mutants within a contribution and hinder assessing the quality of the contributed tests.}
{For instance, in the previous example, \textit{testTriangle5()} and \textit{testTriangle6()} kill no mutant, suggesting they are the most fragile contributed test methods.}
However, the system score is unaffected because their mutants are killed by other tests.
{Thus, the 100\% mutation score neglects quality difference between the test methods, from the mutation analysis perspective.}

\section{Mutation Testing at Method Level}
\label{sec:approach}


\subsection{Test Method Mutation}

\begin{figure}[H]
    \centering
    \includegraphics[width=\linewidth]{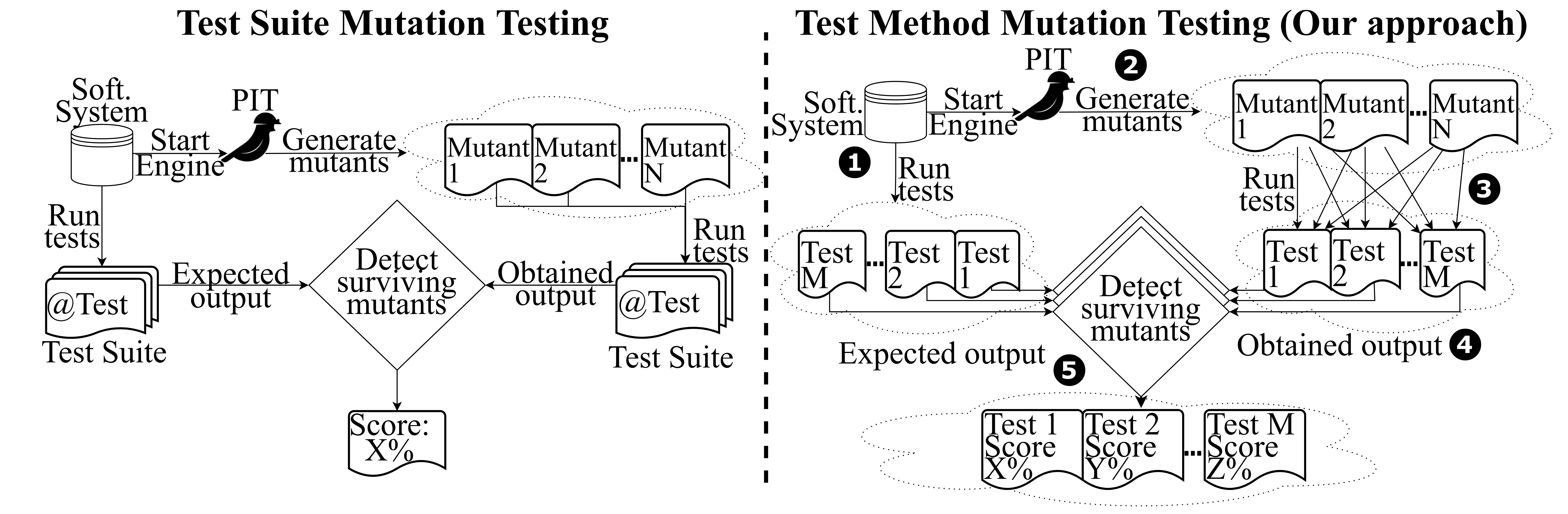}
    
    
    \caption{Traditional approach (left) vs. ours (right).}
    \label{fig:OriginalPITFlow}
    \label{fig:DetailedCSVFlow}
\end{figure}

To address the discussed limitations, we propose a five steps approach, as detailed in Figure~\ref{fig:DetailedCSVFlow}-right.
The first three steps are similar to the traditional mutation testing:
(1) run the test suite and collects the expected output; 
(2) parse the project and apply mutation operators; 
(3) the resulting mutants are separately tested by the test suite; 
(4) the result of \emph{each} executed test method on \emph{each} covered mutant is collected as the obtained output; and
(5) the obtained output is compared to the expected result and the scores are computed for \emph{each} test method.
This approach is implemented by extending the mutation testing tool PIT tool~\cite{pitest} and is publicly available at: \url{https://github.com/victorgveloso/Detailed-CSV-Report-PITest}.



    
    
    
    
    

{The score for a test method \emph{test} is the ratio of mutants killed by the \emph{test} and the total number of mutants the \emph{test} covers.} 
%
%
{Given a test method, its \emph{survived} mutants set is formed by the successful runs and its \emph{killed} mutants by failures and errors.} 
%
{Notice that we do not include the \emph{time-out} set to avoid noise in the collected output, which degrades the ability to define test methods quality.}

\subsection{Example: Computing Test Method Scores}

\begin{figure*}[t]
    \centering
    \includegraphics[width=.9\linewidth]{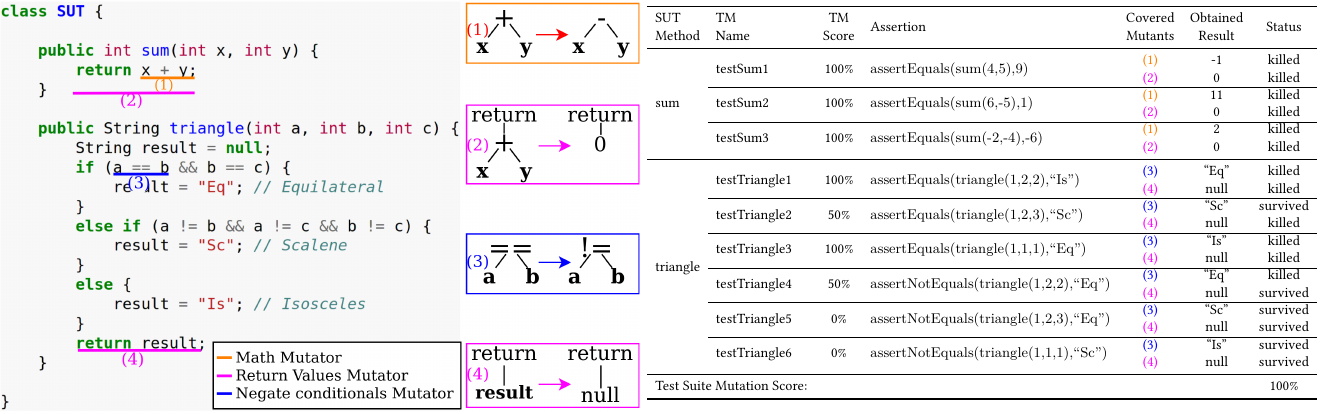}
    \caption{Score computation example in mutation testing inspired by~\cite{fuzzingbook2019_index} (``TM'': Test Method).}
    \label{fig:ExampleTable}
\end{figure*}

{In Figure~\ref{fig:ExampleTable}, we note that 5 out of the 9 test methods have a mutation score of 100\% (column ``TM Score''), two have a score of 50\%, and two have a score of 0\%.}
Both \textit{testTriangle5()} and \textit{testTriangle6()} scores are 0\%, suggesting they have less quality.
Indeed, their assertions (\ie~\textit{assertNotEquals}) are the weakest in the test suite. 

\section{Study Design}
\label{sec:design}

\subsection{Selecting the Software Systems}
{We collect the top 15 Java repositories from GitHub (in terms of the star metric~\cite{icsme2016, DBLP:journals/corr/abs-1811-07643}) and the Apache Commons Lang.}
Next, for each project, we clone the latest master branch version, manually configure the extended PIT~\cite{pitest} via their build configuration file, and discard the projects accused by PIT of having failing tests.
The five remaining projects are highly active and their size ranges from 35.9KLOC (Retrofit) to 310.8KLOC (RxJava).

\subsection{Running the Mutation Testing Tool}\label{sec:mutationtool}

After selecting the target projects, we start the mutation testing execution phase. 
Table~\ref{tab:test-qa} summarizes this analysis: in total, PIT detected 18,321 test cases in the five projects.
Overall, it generated 55,427 mutants, which resulted in 16,149,383 mutant executions.
The mutation scores are overall high, ranging from 73\% (Okhttp) to 86\% (Commons Lang).
For comparison purposes, we also present the coverage values in the last column.
As expected~\cite{fuzzingbook2019_index}, the coverage values are frequently higher than the mutation score. 


\begin{table}[H]
\small
\centering
\caption{Projects test quality overview.}
\label{tab:test-qa}
\begin{tabular}{lrrrrr}
\toprule
\textbf{Project}        & \textbf{Tests} & \textbf{Mutants} & \textbf{TM Runs} & \textbf{Score}  & \textbf{Cov.} \\ \midrule
Commons Lang    &   3,668   & 13,517 & 1.243M    & 86\%  & 95\%  \\ 
RxJava          &   12,145  & 22,342 & 6.368M       & 85\%  & 100\%    \\
ZXing           &   408     & 11,918 & 1.121M       & 75\%  & 94\%     \\
Retrofit        &   337     & 883    & 0.154M       & 75\%  & 51\%     \\
Okhttp          &   1,763   & 6,767  & 7.261M       & 73\%  & 86\%     \\
\bottomrule
\end{tabular}
\end{table}


\subsection{Selecting the Test Methods}

The next step is to select the test methods to be analyzed.
To be selected for this study, test methods must:
(1) contain a \textit{@Test} annotation or a name prefixed by \emph{test},
(2) not rely on anonymous classes,
(3) not contain neither \textit{@Ignore} nor \textit{@Disabled} annotations, and
(4) have a mutation score computable by PIT, \ie it covers at least one mutant.
{Next, we collect the mutation score of the test methods individually, extract the top-100 methods and bottom-100 methods in terms of mutation score, and randomly select 100 methods. }

\subsection{Assessing the Research Questions}

\subsubsection{RQ1 (Quality)}

In the first RQ, we investigate high and low-quality test methods' code and evolution by computing six metrics:
three evolutionary metrics (from PyDriller~\cite{Spadini2018}) and three test-related (from ``tsDetect''~\cite{10.1145/3368089.3417921}).
Those are largely adopted tools in the testing and software mining literature~\cite{catolino2019experience, lenarduzzi2019technical, kazerouni2019assessing, 9240691, 10.1145/3379597.3387453}. 

\noindent\textbf{Test Size}. 
We assess the size of the test methods in terms of \emph{source lines of code} (SLOC).
{Big test methods are heavy and hard to read~\cite{10.1145/3368089.3417921}.}


\noindent\textbf{Test Quality}.
We assess three metrics specific to test methods~\cite{10.1145/3368089.3417921}.
\emph{Number of exceptions} measures the amount of exception-related code structures. 
We compute the \emph{number of bad asserts} (\ie asserts without an explanation) present in test methods~\cite{10.1145/3379597.3387453}.
Lastly, \emph{magic numbers} are direct references to numbers in the tests.


\noindent\textbf{Contributors}.
We assess \emph{developers' expertise} as the ratio of commits they authored in the target project and \emph{number of contributors} is how many distinct developers changed each test method.


\noindent\textbf{Modifications}.
Evolution is an important aspect of any source code and tests are not different.
We analyze the number of changes (commits) in the test methods {to understand their stability.} 


\noindent\underline{Rationale}.
Assessing to what extent high and low-quality test methods are associated with code evolution and static metrics is relevant for both practitioners and researchers. 
Practitioners may consider using static metrics which are cheaper in terms of space and time as a proxy of test quality.
{On the research side, this may support the prediction of test method quality~\cite{8304576} based on both metrics.}

\subsubsection{RQ2 (Test Smells)}
{This RQ assesses the impact of test smells (\ie sub-optimal design choices made when developing tests~\cite{meszaros2007xunit}) on test methods in terms of mutation score.
Like RQ1, we rely on ``tsDetect''~\cite{10.1145/3368089.3417921} and analyze the latest version of the repositories' master branch.
We assess ten test smells~\cite{van2001refactoring, 10.1145/3368089.3417921}}: 
Assertion Roulette,
Duplicate Assert,
Conditional Test Logic,
Dependent Test,
Sleepy Tests,
Sensitive Equality,
General Fixture,
Magic Number Test,
Exception Catching Throwing, and
Unknown Tests.
We select the top 10 most consolidated test smells and discard the ones:
unrelated to test methods (e.g., Constructor Initialization),
debatable in the literature (e.g., Mystery Guest), and
infrequent in our dataset (e.g., Default Test).

\noindent\underline{Rationale}.
Recent studies focus on test smells~\cite{Pecorelli2021}, 
{their impact on defect and change-proneness~\cite{8529832}, and}
co-occurrence with code smells~\cite{10.1145/2970276.2970340}.
{Still, their relationship with mutation score is unclear.}

\section{Results}
\label{sec:results}

\subsection{RQ1 (Quality)}\label{sec:qualityRQ}

Table~\ref{tab:rq1MetricsDescribe} summarizes the metric values for the the best (top-100), random (100-random), and worst (bottom-100) methods.
We apply the Mann-Whitney test at \emph{alpha value} = 0.05 and the Cohen's d effect size between the best and worst test methods (column ``Best vs. Worst''). 
We find a statistically significant difference in all metrics, with at least a very small effect.
Next, we highlight some differences.


\noindent\textbf{Number of lines of code.}
The best test methods are only slightly smaller than the worst ones (9 vs. 10, very small effect size).



\noindent\textbf{Number of bad asserts.}
As most asserts are written without any explanation, this metric can be seen as a proxy of ``number of asserts''.
High-quality test methods have, on average, more asserts (3.7) than low-quality ones (1.6), but the difference is only small.




\noindent\textbf{Number of modifications.}
The best test methods are only slightly less modified than the worst ones (mean 3.3 vs. 3.9, small effect).



\begin{table}[h]
\small
\centering
\small \caption{Metrics overview ($\bar{\eta}$: median; Rnd: Random; N: Negligible; VS: Very Small; S: Small; H: Huge).}
\label{tab:rq1MetricsDescribe}
\begin{tabular}{l c | c | c | cc}
\toprule
\multicolumn{1}{c}{\multirow{2}{*}{Metric}} & Best      & Rnd    & Worst & \multicolumn{2}{c}{Best vs. Worst}  \\
& $\bar{\eta}$ & $\bar{\eta}$ & $\bar{\eta}$ & p-value & effect-size \\ \midrule
N\textsuperscript{\underline{o}} of lines of code & 10 & 10 & 9 & $<0.05$ & VS\\ \midrule
N\textsuperscript{\underline{o}} of bad asserts & 2 & 1 & 1 & $<0.05$ & S\\
N\textsuperscript{\underline{o}} of exceptions & 0 & 0 & 0 & $<0.05$ & S\\
N\textsuperscript{\underline{o}} of magic numbers & 0 & 0 & 0 & $<0.05$ & VS\\ \midrule
N\textsuperscript{\underline{o}} of contributors         & 2 & 2 & 2 & $<0.05$ & VS\\ 
N\textsuperscript{\underline{o}} of modifications         & 3 & 3 & 3 & $<0.05$ & S\\ 
Developer expertise & 0.1 & 0.1 & 0.2 & $<0.05$ & VS\\ \midrule
Score & 0.9 & 0.6 & 0 &  $<0.05$ & H\\ \bottomrule
\end{tabular}
\end{table}

\subsection{RQ2 (Test Smells)}

\begin{figure}[h]
    \centering
    \includegraphics[width=.7\linewidth]{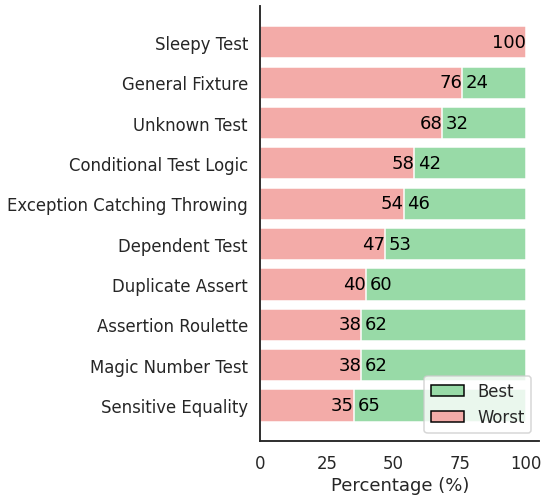}
    \caption{Prevalence of test smells.}
    \label{fig:SmellsStackedBarSortedGrouped}
\end{figure}

{Figure~\ref{fig:SmellsStackedBarSortedGrouped} compares the presence of test smells on both high and low-quality test methods.}
{Sleepy Tests, often related to non-determinism and flaky tests~\cite{10.1145/3338906.3338945, palomba2017does}, only occurs in the worst group.}
{Next, we see that 76\% of General Fixture cases affect low-quality test methods.}
Moreover, 68\% of Unknown Test happen in low-quality test methods.
Finally, Conditional Test Logic and Exception Catching Throwing are also more likely to happen in low-quality test methods, however, the difference is smaller (58\% vs. 42\% and 54\% vs. 46\%, respectively).
On the other hand, we see some test smells occurring more often in the best test methods, \eg Magic Number Test, Assertion Roulette, and Duplicate Assert.
Those test smells are very controversial, for instance, Assertion Roulette represents test methods with more than one assert without explanation/message, which is a common practice in software testing.
Indeed, those test smells are more related to test readability and do not directly affect the ability of the test to catch bugs.

\section{Discussion}
\label{sec:discussion}


\noindent\textbf{Code and evolutionary characteristics.}
It is conventional wisdom that test methods should be small and non-complex to improve their maintainability~\cite{martin2009clean}.
However, we lack empirical data showing the real benefits of having those factors.
We find no major differences between high-quality and low-quality test methods in terms of size, number of asserts, and modifications.
This opens room for novel research to better understand the differences between high and low-quality test methods.

\noindent\textbf{Test smells.}
Recent studies show that test smells may decrease the understandability and maintainability of the test suites~\cite{10.1145/2970276.2970340, bavota2012empirical, 8529832, bavota2015test}, despite practitioners do not perceive test smells as actual problems~\cite{10.1145/2970276.2970340, Pecorelli2021}.
In this study, we find that low-quality test methods are more likely to include critical test smells.
For example, low-quality test methods are over-concentrated on Sleepy Test, General Fixture, and Unknown Test.
On the other hand, high-quality test methods have less critical test smells, which are related to test readability, like Magic Number Test and Assertion Roulette.
Thus, practitioners in charge of maintaining test suites should be aware that the presence of some test smells is associated with the test suite's ability in catching real bugs.

\section{Threats to Validity}
\label{sec:threats}




\noindent\textbf{Timed out tests.} PIT implements heuristics to identify mutants suffering from infinite loops. 
{We discarded \emph{time-out} occurrences from the test method score's formula to prevent noise in the score.}



\noindent\textbf{Failing test suite.} Mutations to static members~\cite{8728964} and tests depending on a specific execution order may be falsely accused of having a non-green test suite. 
{Solved by forcing PIT to execute mutants in separate processes and discarding the failing projects.}


\noindent\textbf{Anonymous classes.} We discard test methods using anonymous classes, due to ``tsDetect'' tool~\cite{10.1145/3368089.3417921} incompatibility.





\noindent\textbf{Generalization.}
We analyzed thousands of test methods provided by open-source Java projects.
However, our findings 
may not be directly generalized to other systems, as commercial ones with closed source and implemented in other languages.


\section{Related Work}
\label{sec:related-work}

Measuring test effectiveness through Mutation Testing is a largely studied topic with well-defined benefits and constraints~\cite{tse_mutation_2011}. 
Addressing those points, Jia~\etal~\cite{tse_mutation_2011} summarize 390 studies in a public repository. 
On the other hand, Giovanni~\etal~\cite{9240623} found that researchers and practitioners perceive existing metrics assist detecting low-quality test suites, but do not guarantee the high quality of a test suite.~\cite{9240623}. 
Catolino~\etal~\cite{catolino2019experience} find assertion density correlation with developer's experience and class-related factors.

Test smells are associated with several factors in software development, for example, code smells~\cite{10.1145/2970276.2970340}, change-proneness and defect-proneness~\cite{8529832}, and post-release defects~\cite{Pecorelli2021}.
Despite the richness of the test smell research topic, practitioners do not perceive test smells as actual problems~\cite{10.1145/2970276.2970340, Pecorelli2021}, 90\% of test smells are never fixed, and fixing takes, on average, 100 days~\cite{10.1145/2970276.2970340}.

Hilton~\etal assess the impact of finer granularity reports on some test coverage limitations. 
The authors describe how non-code changes impact test coverage and how finer granularity reports enable developers to better understand the quality of a specific change~\cite{hilton2018large}.
Despite being considerably cheaper than mutation testing, test coverage still has challenges when adopted in large-scale projects~\cite{8804462}. 
{Challenges aggravated by the monorepo settings at Facebook, where test prioritization usage reduced infrastructure overhead~\cite{8804462}.}

\section{Conclusion}
\label{sec:conclusion}

We proposed an empirical study to assess the quality of \emph{test methods} by relying on mutation testing at the method level.
We show empirical evidence that there are no major differences between high-quality and low-quality test methods in terms of size, number of asserts, and modifications.
Low-quality test methods are over-concentrated on critical test smells, while high-quality test methods are likely to contain less important ones.

    
    
    
    



As future work, we plan to extend our dataset and to include more static and runtime metrics in the analysis to better assess test quality, \eg metrics related to the adopted assertions and the input data.
Lastly, we plan to provide more qualitative analysis on the differences between high and low-quality test methods.

\section*{Acknowledgment}
\label{sec:acknowl}

This research is supported by CAPES, CNPq, and FAPEMIG.


\bibliographystyle{ACM-Reference-Format}
\bibliography{main}



\end{document}